\documentclass[useAMS,usenatbib]{tellus}
\usepackage{graphicx}

\begin{document}

\title[Mechanically-stirred buoyancy-driven circulations]
{The effect of mechanical stirring on horizontal convection}

\author[Author running head]{By R TAILLEUX$^1$\thanks{Corresponding
author.\hfil\break e-mail: R.G.J.Tailleux@reading.ac.uk}
and L ROULEAU} 
\affiliation{$^{1}$Department of Meteorology, University
of Reading, Earley Gate, PO Box 243, Reading, United Kingdom}

\history{Manuscript received 24 November 2003; in final form 25
November 2003}

\maketitle
\begin{abstract}

  The theoretical analysis of the energetics of mechanically-stirred
horizontal convection for a Boussinesq fluid yields the formula:
$$
   G(APE) = \gamma_{mixing} G(KE) + (1+\gamma_{mixing})
   W_{r,laminar}
$$
where $G(APE)$ and $G(KE)$ are the work rate done by the buoyancy 
and mechanical forcing respectively, $\gamma_{mixing}$ is the mixing
efficiency, and $W_{r,laminar}$ is the background rate of increase
in gravitational potential energy due to molecular diffusion. The
formula shows that mechanical stirring can easily induce a very strong
buoyancy-driven overturning cell (meaning a large $G(APE)$) even for
a relatively low mixing efficiency, whereas this is only possible in
absence of mechanical stirring if $\gamma_{mixing} \gg 1$. Moreover,
the buoyancy-driven overturning becomes mechanically controlled when
$\gamma_{mixing} G(KE) \gg (1+\gamma_{mixing}) W_{r,laminar}$.
This result explains why the buoyancy-driven
overturning cell in the laboratory experiments by 
\cite{Whitehead2008} is amplified by the lateral
motions of a stirring rod. The formula implies that the thermodynamic 
efficiency of the ocean heat engine, far from being negligibly small as is
commonly claimed, might in fact be as large as can be thanks
to the stirring done by the wind and tides. These ideas are further
illustrated by means of idealised numerical experiments. A non-Boussinesq
extension of the above formula is also given.

\end{abstract}

\section{Introduction}

  A fundamental objective of ocean circulation theory is to characterise
and quantify the relative importance of the mechanical and thermodynamical
forcing in driving and stirring the oceans. Although the oceans are 
highly nonlinear a system, ocean circulation theory nevertheless historically
evolved under the assumption that separate theories for the wind-driven
and buoyancy-driven circulation could be developed, as it the two kind of
circulation could be regarded as somehow decoupled. From a thermodynamic
viewpoint, this is somehow justified from the fact that mechanical forcing,
unlike the buoyancy forcing, does not modify the fluid parcels' buoyancy,
suggesting that distinct effects should be observable. In the thermodynamic
engineering literature, mechanical forcing is commonly regarded as ``shaft''
work, which alters the energy of the fluid without altering its total 
entropy. In oceanographic textbooks, wind-driven circulation theories
usually occupy a more prominent place that buoyancy-driven circulation
theories. Indeed, it turns out to be relatively easy to link the near
surface transport (Ekman theory) or the vertically-integrated transport
(Sverdrup theory) to the wind stress or its spatial derivatives in a
quantitative relatively accurate way. Theories for the buoyancy-driven
circulation, on the other hand, are often more qualitative, and at best
offer only tentative scaling arguments to relate the strength of the
large-scale overturning cell taking place in the meridional/vertical plane,
e.g. \cite{ACDV1993}. From a physical viewpoint, the two main salient
ingredients of the buoyancy-driven circulation have been traditionally
associated with high-latitude cooling, envisioned as the destabilising
mechanism setting up the meridional overturning circulation into motion,
and turbulent diapycnal mixing, envisioned as the process required to
carry heat from the surface at the depths cooled by deep water formation,
thus precluding the oceans to fill up with dense and salty waters.

\par

  Physically, however, the buoyancy-driven circulation can only be
regarded as independent of the mechanical forcing if one can establish
that the stirring required to maintain turbulent diapycnal mixing is
predominantly sustained by the work rate done by the surface buoyancy
fluxes $G(APE)$ (i.e., the rate of available potential energy
production, see \cite{Lorenz1955}).
A crucial question, therefore, is how large is $G(APE)$?
In their observational analysis of ocean energetics, \cite{Oort1994}
found
 $G(APE)=1.2\pm 0.7 {\rm TW}$, and concluded that the buoyancy forcing
was as important as the wind in forcing and stirring the oceans.
This result, however, was challenged by \cite{Munk1998} 
(MW98 thereafter), who contended that the so-called
\cite{Sandstrom1908}'s theorem forbids $G(APE)$ to be significant,
and hence that turbulent diapycnal mixing must be primarily be sustained
by the work rate done by the mechanical forcing due to the wind and tides,
leading to the idea that the buoyancy-driven circulation is actually
mechanically-driven. MW98 furthermore sought to quantify the magnitude
of the mechanical sources of stirring required to sustain diapycnal mixing
in the oceans. Their main result is the following formula:
\begin{equation}
       G(KE) = \frac{W_{r,forcing}}{\gamma_{mixing}} ,
      \label{MW98_constraint}
\end{equation}
where $G(KE)$ is the work rate done by the mechanical sources of stirring,
$W_{r,forcing}$ is the rate at which the background gravitational potential
energy $GPE_r$ decreases as the result of high-latitude cooling, and
$\gamma_{mixing}$ is the so-called mixing efficiency, e.g.,
\cite{Osborn1980}. MW98 estimated $W_{r,forcing}\approx 0.4\,{\rm TW}$
from assuming the deep water formation rate to be known, and used
the canonical value $\gamma_{mixing}=0.2$ for the mixing efficiency
to conclude that $G(KE)=O(2\,{\rm TW})$ was required to sustain the
observed rate of turbulent diapycnal mixing and its associated poleward
heat transport of about $2\,{\rm PW}$. Since the wind forcing provides
no more than about $1\,{\rm TW}$, this result suggests an apparent
shortfall of about $1\,{\rm TW}$ of mechanical stirring that MW98
argued must only come from the work rate done
by the tides, spawning much research over the past decade on the issue
of tidal mixing.

\par

  MW98's study prompted much debate over the past decade about
the relative importance of the surface buoyancy forcing, and about whether
Sandstrom's theorem really implied for $G(APE)$ to be negligible.
Recently, \cite{Tailleux2009} revisited a number of misconceptions
about the nature of energy conversions in turbulent stratified fluids
by extending the available potential energy framework of \cite{Winters1995}
to the fully compressible Navier-Stokes equations (CNSE thereafter).
One of the main result was to demonstrate from first principles that:
\begin{equation}
      G(APE) \approx W_{r,forcing},
\end{equation}
which physically states that the production rate of $APE$ is approximately
equal to the rate at which $GPE_r$ decreases as the result of high-latitude
cooling. This result is extremely important, because it rigourously proves
that it is inconsistent to assume $W_{r,forcing}$ to be
large while simultaneously assuming $G(APE)$ to be negligible, as done in
MW98. As a result, rather than establishing that the work rate done by
surface buoyancy forcing is small, MW98's study actually establishes precisely
the contrary, since their estimate $W_{r,forcing} \approx 0.4\,{\rm TW}$
implies $G(APE) \approx 0.4\,{\rm TW}$, which turns out to be very close to
\cite{Oort1994}'s lower bound for $G(APE)$. Moreover, \cite{Tailleux2009}
was able to show that a generalisation of MW98's Eq. (\ref{MW98_constraint})
for a non-Boussinesq fluid can be rigourously
derived from first principles by combining the mechanical energy budget for
$APE$ and $GPE_r$ separately, the final result taking the form:
$$
     G(KE) = \frac{1+(1-\xi)\gamma_{mixing}}{\xi \gamma_{mixing}}G(APE) 
$$
\begin{equation}
    = \frac{1-\xi R_f}{\xi R_f} G(APE),
   \label{general_constraint}
\end{equation}
where $\xi$ is a non-Boussinesq nonlinearity parameter which is such that
$-\infty < \xi \le 1$ for water and seawater, $\gamma_{mixing}$ is the 
mixing efficiency, and $R_f=\gamma_{mixing}/(1+\gamma_{mixing})$ is the 
so-called flux Richardson number, e.g., \cite{Osborn1980}. MW98's 
Eq. (\ref{MW98_constraint}) is recovered in the limiting case $\xi=1$
describing a Boussinesq fluid with a linear equation of state, by using
the above result that $G(APE)\approx W_{r,forcing}$.

\par

 The fact that the generalisation of MW98's result can be written as
a formula linking $G(KE)$ and $G(APE)$ rather than $G(KE)$ and
$W_{r,forcing}$ raises the question of whether it is really appropriate
to interpret Eq. (\ref{general_constraint}) (and hence Eq. 
(\ref{MW98_constraint})) as a constraint on the amount of $G(KE)$ required
to sustain diapycnal mixing in the oceans, as suggested by MW98. Indeed, 
it seems that such an interpretation implicitly assumes that both $G(APE)$
and $\gamma_{mixing}$ can be regarded as fixed in some sense, but whether
this is the correct way to interpret
Eq. (\ref{general_constraint}) is unclear, given that the latter can
be rewritten in the two following ways: 
\begin{equation}
     G(APE) = \frac{\xi \gamma_{mixing}}{1+(1-\xi)\gamma_{mixing}} G(KE),
     \label{gape_constraint}
\end{equation}
or as:
\begin{equation}
    \xi R_f = \frac{G(APE)}{G(KE)+G(APE)} ,
    \label{xirf_constraint}
\end{equation}
i.e., under the form of a constraint either on $G(APE)$ or on
$\xi R_f$. Note that from observations, plausible numbers for the
wind and buoyancy forcing are $G(KE) \approx 1\,{\rm TW}$ and 
$G(APE) \approx 0.5\,{\rm TW}$, which, if inserted into Eq. 
(\ref{xirf_constraint}) yields:
$$
    \xi R_f \approx \frac{0.5}{1.5} = 0.3333 .
$$
Another way to pose the problem initially formulated by MW98 would
be to ask the question of whether the above value for $\xi R_f$ can
be ruled out from what we know about mixing efficiency (and the role
played by the nonlinearity of the equation of state, but we know less
about the latter). Although the value $\gamma_{mixing}\approx R_f
\approx 0.2$ is often assumed, it is important to realize that such
value mostly pertains to mechanically-driven turbulent mixing, as 
occurs in relation with shear flow instability for instance.
Indeed, mixing efficiency values as high as $R_f=0.5$ ($\gamma_{mixing}=1$)
have been reported by \cite{Dalziel2008} for buoyancy-driven mixing
associated with Rayleigh-Taylor instability. Since $G(APE)$ is 
comparable to $G(KE)$, values of $\gamma_{mixing}$ intermediate between
those for mechanically- and buoyancy-driven mixing should be expected.

\par

  In trying to figure out the exact physical meaning of 
Eq. (\ref{general_constraint}), it is important to discuss the nature and
structure of $G(KE)$ and $G(APE)$ in more details. Thus, the work rate done 
by the wind stress is given by:
\begin{equation}
      G(KE) = \int_S {\bf \tau}\cdot {\bf u}_s dS
      \label{wind_work}
\end{equation}
where $\tau$ is the wind stress at the ocean surface, and ${\bf u}_s$
is the ocean surface velocity. The important point about $G(KE)$ is that
it is a correlation between the external forcing (the wind stress) and
a parameter depending upon the particular state of the system (the surface
velocity). In other words, it is essential to realize that $G(KE)$ cannot
be determined from the knowledge of the external forcing alone. 
This is important, because Eq. (\ref{wind_work}) expresses the possibility
for the wind-driven circulation to be controlled by the buoyancy forcing, 
to the extent that the latter is able to influence the surface velocity.
In order to gain insight into the potential importance of the buoyancy 
control of the wind-driven circulation, it would be interesting to compute
$G(KE)$ for a purely wind forced homogeneous ocean model.
Doing so, however, is beyond the scope of this paper. For the present 
purposes, we shall assume that $G(KE)$ is primarily determined by the wind
forcing, and that buoyancy forcing only alters it as a second order effect,
which seems plausible on the basis that the near surface circulation in the
oceans is usually thought to be the signature of the wind forcing rather 
than the buoyancy forcing.

\par

 With regard to the work rate done by surface buoyancy fluxes, it
was shown by \cite{Tailleux2009} to be given by the following 
expression:
\begin{equation}
   G(APE) = \int_{S} \frac{T-T_r}{T} Q_{surf} dS ,
   \label{GAPE_exact}
\end{equation}
in the case of a compressible thermally-stratified fluid forced by
surface heat fluxes,
where $T$ is the surface temperature of the fluid parcels, and
$T_r$ the temperature the surface parcels would have if displaced
adiabatically to their level in \cite{Lorenz1955}'s reference state,
while $Q_{surf}$ is the diabatic rate of heating cooling/heating due
to the surface heat fluxes. Useful approximations for $G(APE)$ can
be obtained by expanding $T$ as a Taylor series expansion around
the surface pressure $P_a$, i.e., $T \approx T_r + \Gamma_r (P_a - P_r) + 
O((P_a-P_r)^2)$, where
$\Gamma_r = \alpha_r T_r/(\rho_r C_{pr})$ is the adiabatic lapse
rate, leading to:
$$
     G(APE) \approx -\int_{S} \frac{\alpha_r (P_r-P_a)}{\rho_r C_{pr}}
      Q_{surf} dS
$$
\begin{equation}
        \approx \int_{S} \frac{\alpha_r g z_r}{C_{pr}} Q_{surf} dS ,
     \label{GAPE_formula}
\end{equation}
by using the approximation $P_r-P_a \approx -\rho_0 g z_r$, the latter
expression being actually the Boussinesq approximation of $G(APE)$
derived by \cite{Winters1995} provided that one regards the 
thermal expansion $\alpha_r$ and specific heat capacity $C_{pr}$ as
constant (the suffix $r$ means that the variables have to be estimated
in \cite{Lorenz1955}'s reference state). As for $G(KE)$, $G(APE)$ also
takes the form of a correlation between the external forcing (the 
heating/cooling rate at the surface) and a parameter depending on the
particular state of the system, namely 
$(T-T_r)/T \approx \alpha_r (P_a-P_r)/(\rho_r C_{pr})
\approx \alpha_r g z_r/C_{pr}$. Eqs. (\ref{GAPE_exact}) and
(\ref{GAPE_formula}) make it possible, therefore, for the buoyancy-driven
circulation to be mechanically-controlled. Physically, this is expected,
because mechanical forcing is widely agreed to increase diapycnal mixing,
which should produce a deeper thermocline, thereby allowing dense plumes
to penetrate deeper, thus increasing $G(APE)$ and hence the buoyancy-driven
circulation. Interestingly, this appears consistent with the laboratory
experiments by \cite{Whitehead2008} which provide evidence that the lateral
motion of a stirring rod can greatly enhance the strength of horizontal
convection (see \cite{Hughes2008} for a review on this topic).
This paper examines the possibility of interpreting this result as the
consequence that $G(APE)$, and hence the buoyancy-driven circulation,
is increased by the work rate done by the stirring rod. 
 To that end, Section 2 seeks to extend Eq. (\ref{general_constraint}) to 
also describe the purely buoyancy-driven case for which it is not valid.
The physical meaning of the extended formula is then explored by means
of idealised numerical experiments in Section 3. 
Finally, section 4 presents a discussion of the results.


\section{Theoretical results about the effects of mechanical stirring
on the work rate done by surface buoyancy fluxes}

\subsection{Summary of Tailleux (2009)'s theory}

 Our starting point is the theoretical description of the energetics of
mechanically and thermodynamically forced turbulent stratified fluids 
recently derived by \cite{Tailleux2009} pertaining to a thermally-stratified
fluid governed by the fully compressible Navier-Stokes equations (NCSE
thereafter), which builds upon the available potential energy framework
previously introduced by \cite{Winters1995} for a Boussinesq fluid with
a linear equation of state. In \cite{Tailleux2009}'s description, the
energetics is described at leading order by means of the following five
evolution equations:
\begin{equation}
  \frac{d(KE)}{dt} = -C(KE,APE) + G(KE) - D(KE) ,
  \label{KE}
\end{equation}
\begin{equation}
  \frac{d(APE)}{dt} = C(KE,APE) + G(APE) - D(APE) ,
  \label{APE}
\end{equation}
\begin{equation}
  \frac{d(GPE_r)}{dt} = W_{r,turbulent} + W_{r,laminar} - W_{r,forcing} ,
  \label{GPEr}
\end{equation}
\begin{equation}
  \frac{d(IE_0)}{dt} = (1-\Upsilon_0) \dot{Q}_{net} + D(APE) + D(KE) - G(APE) ,
  \label{IEo}
\end{equation}
\begin{equation}
  \frac{d(IE_{exergy})}{dt} = \Upsilon_0 \dot{Q}_{net} - W_{r,turbulent}
   - W_{laminar} ,
  \label{IEex}
\end{equation}
where $KE$ is the volume-integrated kinetic energy, 
$APE$ is \cite{Lorenz1955}'s volume-integrated available potential energy, 
$GPE_r$ is the volume-integrated gravitational potential
energy of \cite{Lorenz1955}'s reference-state, 
$IE_0$ is the volume-integrated of a subcomponent of internal energy $(IE)$
that we call the dead part of $IE$, $IE_{exergy}$ is the volume-integrated
of another subcomponent of $IE$ called the exergy. Physically, variations in
$IE_0$ are associated with variations in the equivalent thermodynamic
equilibrium temperature $T_0(t)$ of the system, whereas variations in 
$IE_{exergy}$ reflect variations in the reference temperature profile
$T_r(z,t)$. The other important conversion terms are $C(KE,APE)$, the 
so-called buoyancy flux that represents the reversible conversion between
$KE$ and $APE$; $G(APE)$, the production rate of $APE$ that physically
represents a conversion between $IE_0$ and $APE$; $W_{r,laminar}$ and
$W_{r,turbulent}$ represent the laminar and turbulent rate of exchange
between $IE_{exergy}$ and $GPE_r$ due to molecular diffusion; 
$D(KE)$ is the dissipation rate of
$KE$ into internal energy $IE_0$ by molecular viscous processes; 
$D(APE)$ is the dissipation rate of $APE$ into internal energy $IE_0$ by
molecular diffusive processes. The parameter $\Upsilon_0 \ll 1$ plays the
role of a thermodynamic efficiency that is very small for a nearly 
incompressible fluid, and which controls how much of the net surface 
heating rate $\dot{Q}_{net}$ splits between $IE_{exergy}$ and $IE_0$.

\par

\begin{figure*}
\centering
\includegraphics[width=10cm]{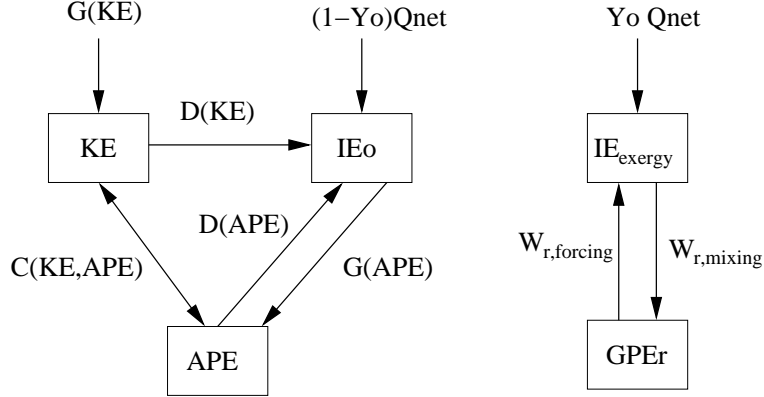}
\caption{Schematic diagram of the energy conversions taking place in a
mechanically and thermodynamically forced turbulent thermally-stratified
fluid corresponding to Eqs. (\ref{KE}-\ref{IEex}) previously derived
by Tailleux (2009).}
\label{gper_budget}
\end{figure*}

 As previously shown by \cite{Winters1995}, a number of conversion terms
appear to be strongly correlated to each other. This is the case for
$D(APE)$ and $W_{r,turbulent}$, owing to both terms:
1) being controlled by molecular diffusion;
2) being controlled by the spectral distribution of the $APE$ density,
see \cite{Holliday1981} and
\cite{Roullet2009} for a discussion of the latter concept.
For a compressible thermally-stratified fluid, \cite{Tailleux2009} showed
that this correlation between $D(APE)$ and $W_{r,turbulent}$ can be written
under the general form:
\begin{equation}
      W_{r,turbulent} = \xi D(APE)
      \label{wr_turbulent}
\end{equation}
where $\xi$ is a parameter that measures the importance of the nonlinearity
of the equation of state, such that $\xi=1$ for a Boussinesq fluid with
a linear equation of state, but $-\infty < \xi \le 1$ for water or 
seawater. The second type of conversion terms that are correlated to
each other are $G(APE)$ and $W_{r,forcing}$ owing to both terms being
controlled by the surface buoyancy forcing. \cite{Tailleux2009} shows 
that to a good approximation, one usually has:
\begin{equation}
     G(APE) \approx W_{r,forcing}.
     \label{GAPE_Wrforcing}
\end{equation}
For the sake of brevity, the reader is referred to \cite{Tailleux2009}
for the details about the explicit forms of all the terms entering the
above energy equations, as they are unimportant for the arguments
developed in this paper. 
Fig. (\ref{gper_budget}) schematically illustrates the energetics
of a mechanically and thermodynamically forced stratified fluids 
associated with the above equations. For all practical purposes,
$G(APE)$ and $G(KE)$ represent the work rate done by the buoyancy
and mechanical forcing respectively, whereas $D(APE)$ and $D(KE)$
represent the two terms dissipating the ``available'' mechanical 
energy $ME=APE+KE$.

\subsection{A general theory for $G(APE)$}

  A fundamental question posed by this paper is what controls the
magnitude of the buoyancy-driven circulation, that is, the circulation
drawing its energy from $G(APE)$, with and without mechanical stirring
acting on the fluid. In order to answer this question, we first need 
to introduce a couple of parameters that are traditionally
used to measure the efficiency of turbulent mixing, the so-called
mixing efficiency $\gamma_{mixing}$ and the flux Richardson number
$R_f$, which are defined here as follows:
\begin{equation}
     \gamma_{mixing} = \frac{D(APE)}{D(KE)} ,
     \label{gamma_def}
\end{equation}
\begin{equation}
 R_f = \frac{\gamma_{mixing}}{1+\gamma_{mixing}}  
= \frac{D(APE)}{D(APE)+D(KE)} .
   \label{rf_def}
\end{equation}
The physical rationale for such definitions, as well as 
their connection to other existing definitions, was discussed
in details in \cite{Tailleux2009} to which the reader is 
referred to for details. From a practical viewpoint, the 
differences in existing definitions is unimportant, as least
in the context of a Boussinesq fluid, as then all definitions
are then numerically equivalent, even if based on different
physical assumptions. In particular, the present definitions
allow to recover MW98's Eq. (\ref{MW98_constraint}), as 
discussed below. To distinguish them from existing definitions,
\cite{Tailleux2009} refer to the above $\gamma_{mixing}$ and
$R_f$ as the ''dissipative'' mixing efficiency and flux
Richardson number respectively.

\par

 In a second step, the mechanical energy balance is constructed
by summing the steady-state version of the $KE$ and $APE$ equations,
which yields:
\begin{equation}
    G(APE) + G(KE) = D(APE) + D(KE) .
\end{equation}
Now, by combining this equation with the
definition of the flux Richardson number (i.e.,
Eq. (\ref{rf_def}), one may write:
\begin{equation}
      D(APE) = R_f [ G(APE) + G(KE) ] .
\end{equation}
Next, we turn to the steady-state $GPE_r$ balance, viz.,
\begin{equation}
     W_{r,turbulent}+W_{r,laminar} = W_{r,forcing} .
\end{equation}
As mentioned above, one has to a very good approximation
the following equality $W_{r,forcing} \approx G(APE)$. 
For mathematical rigour, we introduce a parameter 
$\xi_2 \approx 1$ making exact the equality
$W_{r,forcing}=\xi_2 G(APE)$.
Using the definition of $\xi$ above, the $GPE_r$ budget
provides the following relationship:
\begin{equation}
    \xi D(APE) + W_{r,laminar} = \xi_2 G(APE),
\end{equation}
which after some algebra eventually yields:
\begin{equation}
   G(APE) = \frac{\xi R_f}{\xi_2 - \xi R_f} G(KE) 
   + \frac{W_{r,laminar}}{\xi_2 - \xi R_f} . 
  \label{general_constraint2}
\end{equation}
Eq. (\ref{general_constraint2}) is one of the most 
important result of this paper, as it represents one
way to demonstrate the interconnection of the work
rate done by the wind and buoyancy forcing in general.
A useful limit is the case of the widely used 
Boussinesq model for which $\xi=\xi_2 = 1$, in which
case Eq. (\ref{general_constraint2}) simplifies to:
$$
     G(APE) = \frac{R_f}{1-R_f}G(KE) + \frac{W_{r,laminar}}{1-R_f}
$$
\begin{equation}
    = \gamma_{mixing} G(KE) + (1+\gamma_{mixing})W_{r,laminar} .
    \label{general_constraint_boussinesq}
\end{equation}
Eq. (\ref{general_constraint2}) shows that in absence of 
mechanical forcing, the work rate done by the surface buoyancy
fluxes is given by:
\begin{equation}
     G(APE) = \frac{W_{r,laminar}}{\xi_2 - \xi R_f}
\end{equation}
or, for a Boussinesq fluid:
\begin{equation}
     G(APE) = \frac{W_{r,laminar}}{1-R_f} = (1+\gamma_{mixing})
     W_{r,laminar} . 
\end{equation}
The latter two results are interesting, because they show that
although $W_{r,laminar}$ is small by construction, it 
does not forbid in principle $G(APE)$ to be large provided
that $\xi_2 - \xi R_f$ or $1-R_f$ can become small enough.
The problem, however, is that the latter conditions require
$R_f$ to be close to unity for a Boussinesq fluid, or 
equivalently $\gamma_{mixing} \gg 1$, which is much larger
than the widely used canonical value $\gamma_{mixing}
\approx 0.2$. On the other hand, we are not aware of any
published value for $\gamma_{mixing}$ pertaining to horizontal
convection, so that sufficient empirical evidence to be really
conclusive. The idealised numerical experiments presented below
seeks to get insight into this issue.
 
\par

  Perhaps the most important feature of 
Eq. (\ref{general_constraint2}) is to reveal that even
a small amount of mechanical forcing is sufficient to 
radically alter the nature of horizontal convection, as
it suggests that the latter becomes mechanically
controlled when $G(KE)$ is such that 
$\xi R_f G(KE) \gg W_{r,laminar}$, which is possible even
for relatively low values of $\gamma_{mixing}$ --- a very
important point. This theoretical result
suggests therefore that $G(APE)$ can be dramatically
enhanced by the presence of mechanical stirring, provided
that the latter provides
positive work to the system (i.e., such that $G(KE)>0$).
Negative work, on the other hand, should reduce the strength
of the buoyancy-driven circulation. The numerical examples
studied next seek to illustrate these different ideas.

\begin{figure*}
\centering
\includegraphics[width=14cm]{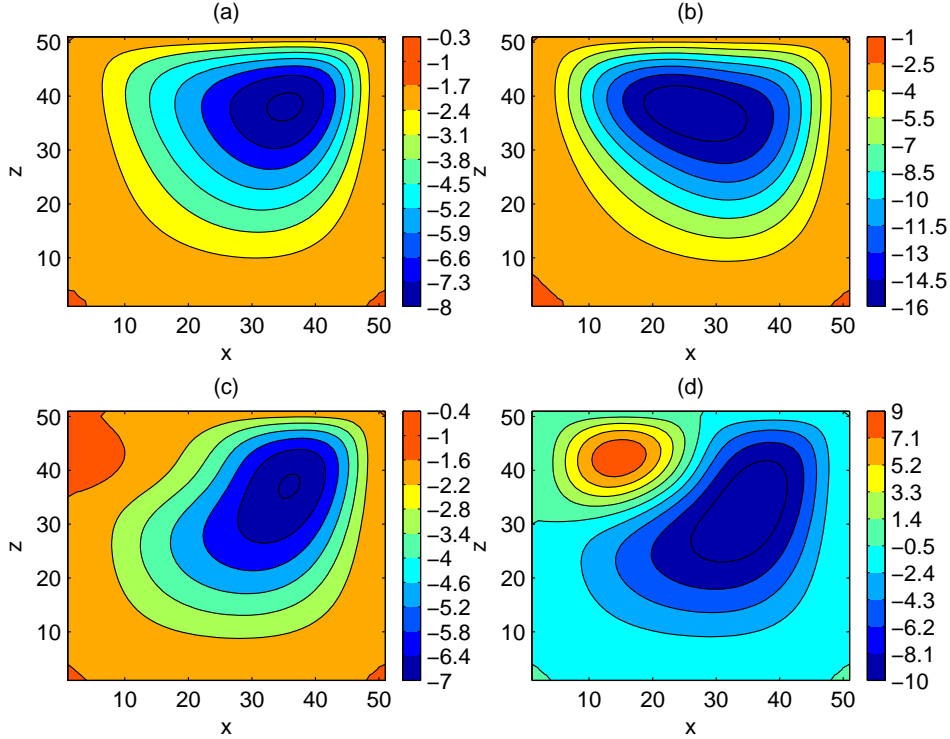}
\caption{Streamfunction for the different experiments: (a)
No mechanical forcing; (b) Clockwise mechanical forcing; (c)
Weak anti-clockwise mechanical forcing; (d) Strong anti-clockwise 
mechanical forcing.
}
\label{4psi}
\end{figure*}

\begin{figure*}
\centering
\includegraphics[width=14cm]{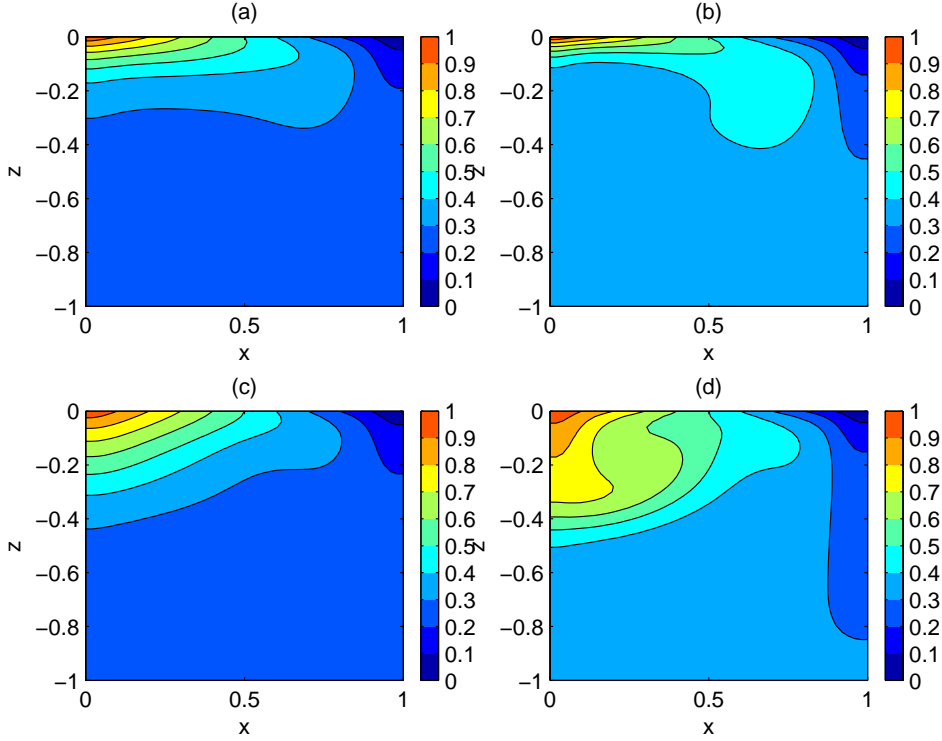}
\caption{Temperature field for the different experiments: (a)
No mechanical forcing; (b) Clockwise mechanical forcing; (c)
Weak anti-clockwise mechanical forcing; (d) Strong anti-clockwise
mechanical forcing.}
\label{4T}
\end{figure*}

\section{Numerical experiments}

  The physical implications of the formula given by Eqs.
(\ref{general_constraint2}) and (\ref{general_constraint_boussinesq})
are explored in the following by means of numerical experiments aiming
at illustrating the main effects mechanical stirring can have on 
buoyancy-driven circulations. 

\subsection{Model description}

  To that end, the idealised problem of horizontal convection previously
considered by \cite{Paparella2002} is investigated here, the novelty being
the addition of mechanical stirring to the problem modelled here as a
forcing term in the vorticity equation. The numerical implementation of
such a model is that previously described by \cite{Marchal2007}. 
The equations solved
by the numerical model are the Boussinesq equations for a fluid with
a linear equation of state:
\begin{equation}
   \frac{\partial \omega}{\partial t} = J(\Psi,\omega)
   - g \alpha \frac{\partial T}{\partial x} + \nu \nabla^2 \omega
   + F(x,z,t)
\end{equation}
\begin{equation}
   \frac{\partial T}{\partial t} = J(\Psi,T) + \kappa \nabla^2 T , 
\end{equation}
where $\omega = \nabla^2 \Psi$ is the vorticity, $T$ is the temperature,
$J(a,b)=\partial a/\partial x \partial b/\partial z-
\partial a/\partial z \partial b/\partial x$ is the Jacobian operator,
$g$ is the acceleration due to gravity, $\alpha$ is the thermal expansion
coefficient assumed to be constant, $\nu$ is the kinematic viscosity,
$\kappa$ is the thermal diffusivity, and $F$ is a term aimed at modelling
the effect of mechanical stirring.

\par

 Following \cite{Marchal2007}, the above system is made dimensionless
as follows: $t=(L^2/\kappa)t^{\ast}$, $(x,z)=L(x_{\ast},z_{\ast})$,
$\omega = (\kappa/L^2)\omega_{\ast}$, $T=\Delta T T_{\ast}$, and
$\Psi = \kappa \Psi_{\ast}$, where the starred quantities are the
dimensionless ones. The dimensionless forms of the above equations
become, after dropping the start for clarity:
\begin{equation}
     \frac{\partial \omega}{\partial t} = J(\Psi,\omega) + 
     P_r \left ( -R_a \frac{\partial T}{\partial x} + \nabla^2 \omega 
    \right ) + F^{\ast},
\end{equation}
\begin{equation}
     \frac{\partial T}{\partial t} = J ( \Psi, T ) + \nabla^2 T ,
\end{equation}
where $R_a = g \alpha \Delta T^3/(\nu \kappa)$ is the Rayleigh number,
and $P_r = \nu/\kappa$ is the Prandtl number. In the numerical experiments
described here, we used $P_r = 10$ and $R_a = 10^5$. For comparison, note
that typical oceanic values are $R_a=O(10^{20})$, e.g.,
\cite{Paparella2002}.

\par

 As in \cite{Paparella2002}, the fluid is forced by a surface temperature
boundary condition varying linearly in $x$.
The numerical resolution used for the experiments is
$51\times 51$ in a 2-D square geometry. At equilibrium, the forcing
term in the vorticity equation is associated with the work rate:
\begin{equation}
     G(KE) = \int \int_{V} F\Psi \,dx dz
\end{equation}
which can be in principle either positive or negative.

\subsection{Experiments}

Four idealised experiments were considered:
\begin{enumerate}
\item Purely buoyancy-driven (i.e., no mechanical forcing);
\item Clockwise Mechanical forcing;
\item Weak anti-clockwise mechanical forcing;
\item Strong anti-clockwise mechanical forcing.
\end{enumerate}

The purely buoyancy-driven case is associated with a clock-wise 
thermally-direct circulation, and represents the ``control'' reference
simulation to be compared with the mechanically-stirred ones.
The purpose of Experiment (ii) is to gain insight into the case for
which the mechanical stirring tends to re-enforce the existing 
thermally-direct clockwise circulation, implying $G(KE)>0$. From Eq.
(\ref{general_constraint_boussinesq}), the expectation is that the
overturning circulation should increase both as the result of increased
$G(APE)$, as well as from the direct effect of the mechanical forcing
acting as a source of clockwise vorticity. In such a case, therefore,
$KE$, $APE$, $\Psi$, $D(APE)$ and $G(APE)$ all should increase.

\par

 In experiments (iii) and (iv), however, the anti-clockwise forcing 
works against the existing buoyancy-driven circulation. If too weak,
as designed to be the case for (iii), the mechanical forcing is 
unable to reverse the sense of the existing thermally-direct circulation
anywhere, resulting in $G(KE)<0$. According to Eq. 
(\ref{general_constraint_boussinesq}), a decrease in all quantities
$KE$, $APE$, $\Psi$, $D(APE)$, and $G(APE)$ is expected in that case.
If the clockwise forcing is strong enough, however, it can locally 
generate a anti-clockwise circulation associated with $G(KE)>0$, making
it possible in that case for all above quantities to be increased.

\subsection{Results}

\begin{table}

\begin{tabular}{ccccc}

   & G(KE) & G(APE) & D(APE) & $\gamma_{mixing}$  \\
   \hline 
(a) & 0    &  5.98       & 4.12       &  1.78          \\
(b) & 7.54    &  10.1       &  9.52      &  1.04                 \\
(c) & -0.77    & 4.72        &  2.58      &   1.49          \\
(d) & 9.83    &  8.50       & 7.18       &  0.60            \\
   \hline
\end{tabular}
\label{table1}
\caption{Values of the work rate done by the mechanical forcing $G(KE)$,
rate of $APE$ production $G(APE)$, diffusive $APE$ dissipation rate $D(APE)$,
and mixing efficiency $\gamma_{mixing}$ for 
(a) No mechanical forcing; (b) Clockwise mechanical forcing; (c)
Weak anti-clockwise mechanical forcing; (d) Strong anti-clockwise mechanical
forcing.}
\end{table}

\begin{table}

\begin{tabular}{ccccc}
     & KE &  APE & $\Psi_{max}$ & $H_{max}$ \\
    \hline
(a)  & 0.13    & 4.04     &  7.2    & 3.15     \\
(b)  & 0.64    & 6.25     &  15.5   &  5.09     \\
(c)  & 0.09    & 5.49     &  6.08   &  2.19   \\
(d)  & 0.51    & 11.1     &  9.67   &  2.75   \\
\hline
\end{tabular}
\label{table2}
\caption{Values of the volume-integrated kinetic energy $KE$, 
volume-integral available potential energy $APE$, maximum value
of the thermally-direct overturning streamfunction, and maximum
value of the heat transport for: (a) no mechanical forcing; (b)
Clockwise mechanical forcing; (c) Weak anti-clockwise mechanical forcing;
(d) Strong anti-clockwise mechanical forcing.}
\end{table}

The results are illustrated by the 
figs. (\ref{4psi}) and (\ref{4T}), which show the streamfunction and
temperature distributions for the 4 different experiments considered. 
In addition, Tables 1 and 2 summarise how
several important quantities vary as a function of the mechanical forcing
imposed. As expected, all the plots for $\Psi$ show evidence of a well-marked
thermally-direct cell in all cases.
Fig. (\ref{4psi}) (b) shows that the strength of the overturning cell is
greatly enhanced by the addition of clockwise mechanical forcing. This can
be explained by the observed increase in $G(APE)$ (Table 1), as well as by
the direct effect of the mechanical forcing as a source of clockwise 
vorticity. Fig. \ref{4T} (b) shows that the increase in $G(APE)$ is 
associated with the dense plumes penetrating  
deeper in that case. Fig. (\ref{4psi}) (c) shows that a weak 
anti-clockwise mechanical forcing decreases the buoyancy-driven circulation,
which is expected from the theory when $G(KE)<0$ as is the case here
(Table 1). The weakening of the circulation results from the decrease in
$G(APE)$, despite an apparent increase in the thermocline depth 
(Fig. \ref{4T}) which one would normally associate with an increase in
$G(APE)$. As shown in Table 2, this occurs because of a reduction in
the strength of the surface heat flux, which results in a decrease in
the maximum heat transport $H_{max}$. 
Fig. (\ref{4psi}) (d) shows that the anti-clockwise
mechanical forcing is strong enough to generate a thermally-indirect
circulation, resulting in $G(KE)>0$, which increases the overturning
strength. In that case, there is no ambiguity that the increase in 
the latter is entirely explained by the increase in $G(APE)$, given that
the mechanical forcing acts as a source of anti-clockwise vorticity,
which can only tend to weaken the buoyancy-driven thermally-direct cell.

\par

 In the literature, there is a tendency to regard the mixing efficiency
$\gamma_{mixing}$ as some kind of universal parameter whose value is 
relatively constant and uniform throughout the oceans. It is interesting,
therefore, to examine whether this is the case in the present numerical
experiments. To that end, the viscous dissipation $D(KE)$ and diffusive
dissipation $D(APE)$ were estimated for all experiments, the values being
reported in Table 1. As it turns out, $\gamma_{mixing}$ is found to vary
widely across the different experiments. Its maximum value is achieved in
the purely buoyancy-driven case, for which $\gamma_{mixing} \approx 1.78$.
Interestingly, this value is significantly larger than the value
$\gamma_{mixing} \approx 1$ reported
by \cite{Dalziel2008} in the context of Rayleigh-Taylor instability,
and which represents so far the highest value of mixing efficiency ever
reported. The Rayleigh-Taylor instability, however, is known to have the
peculiar property that the $APE$ driving the instability is not entirely
available for turbulent mixing. For this reason, \cite{Tailleux2009} 
suggested that higher values of mixing efficiency could in principle be
achieved for buoyancy-driven turbulent mixing events if the above 
constraint could be relaxed. As it turns out, horizontal convection does
not suffer from the limitations attached to the Rayleigh-Taylor instability,
making it possible to achieve values of $\gamma_{mixing}$ significantly
large than unity. The addition of mechanical stirring, however, is seen 
in Table 1 to systematically reduce $\gamma_{mixing}$, the largest reduction
occurring for the strong anti-clockwise mechanical forcing.
In that case, a value of
$\gamma_{mixing}=0.6$ is reached, which is much closer to the value
$\gamma_{mixing}=0.2$ commonly used in the literature. The present numerical
experiments therefore suggest that the mixing efficiency of a fluid forced
both mechanically and thermodynamically is likely to be potentially highly
sensitive to the particular configuration considered. This challenges the
notion that $\gamma_{mixing}=0.2$ should be used in the oceans, given that
$G(APE)$ and $G(KE)$ are comparable in magnitude in reality.

\par

 Another interesting question is what is the effect of the mechanical stirring
on the heat transport, which was also raised in MW98's study.
 The last column of Table 2 provides insight into this issue
by providing the maximum value of the heat transport for each experiment.
The most striking result is that the heat transport appears to be 
drastically increased when the mechanical forcing supports the 
thermally-direct circulation, but decreased when the mechanical forcing
opposes the latter.
 In the actual oceans, the wind forcing creates 
surface Ekman cells that are alternatively helping or opposing the
large-scale thermally direct cell in the Atlantic ocean responsible for
the Atlantic heat transport. It seems therefore difficult to conclude at
this stage whether the overall net effect of the mechanical forcing in
the oceans is to directly contribute to the strength of the overturning
circulation by being a net source of vorticity of the right sign, or if
its contribution is only indirect and limited to its increase of $G(APE)$ and
hence of the strength of the buoyancy-driven component of the overturning.

\section{Discussion and conclusions}

  In this paper, the effect of mechanical stirring on buoyancy-driven flows
  was investigated both theoretically and numerically. The most important
  theoretical result are the formula Eqs. (\ref{general_constraint2}) and
  (\ref{general_constraint_boussinesq}) linking the work done by the 
  mechanical stirring to the work rate done by the surface buoyancy fluxes
  via the bulk mixing efficiency of the system, valid for the compressible
  Navier-Stokes equations and for the Boussinesq model respectively.
  These formula are a further generalisation of that derived by 
  \cite{Tailleux2009}, which extends \cite{Munk1998}'s previous result
  on the energy requirement on the mechanical sources of stirring to 
  sustain diapycnal mixing in the oceans. The actual physical meaning of
  the formula Eqs. (\ref{general_constraint2}) and 
  (\ref{general_constraint_boussinesq}) is not entirely clear, however.
  Physically, the formula only seem to express the existence of a 
  mechanical control on buoyancy-driven flows and vice-versa, rather 
  than a true constraint on $G(KE)$, in contrast with the interpretation
  put forward by \cite{Munk1998}. This idea is further re-enforced by the 
  structure of $G(KE)$ and $G(APE)$, which demonstrate that the work 
  rate done by the mechanical and buoyancy forcing does not simply depend
  on the external forcing (i.e., the wind and surface buoyancy fluxes), 
  but also on the actual state of the system. In other words, the work 
  rate done by either the wind and buoyancy forcing depends sensitively
  on the work rate done by the other kind of forcing. In this paper,
  this idea is examined from the viewpoint of the buoyancy-driven 
  circulation, by demonstrating that the presence of mechanical stirring
  can have dramatic effects on the value of $G(APE)$ owing to its direct
  effect on diapycnal mixing and hence on the structure of the thermocline.
The other way that mechanical forcing can affect $G(APE)$ is by modifying
the near surface temperature, thereby modifying the net flux into the system,
since the flux is also to be determined as part of the solution when
a surface temperature boundary condition is imposed.

\par

 In the purely buoyancy-driven case, Eq. (\ref{general_constraint2})
demonstrates that large values of $G(APE)$ are in principle achievable
provided that the ``dissipative'' mixing efficiency 
$\gamma_{mixing}=D(APE)/D(KE)$ defined in \cite{Tailleux2009} is large 
enough. The particular numerical experiment considered in this paper 
suggests that this is possible, since the value $\gamma_{mixing}\approx 2$
was reached in the purely buoyancy-driven case, which is one order of
magnitude larger than the value $\gamma_{mixing}=0.2$ currently assumed
in most studies of turbulent mixing. Although the result is only tentative,
since a more systematic investigation is required to ascertain the 
robustness of the results, it is important to note that this appears to
be consistent with the numerical results of \cite{Paparella2002} that
show entropy production in horizontal convection
to increase with the Rayleigh number. According to our theoretical results,
this is indeed only possible if $D(APE)$ and hence $G(APE)$ also increase
with the Rayleigh number. On the other hand, \cite{Paparella2002}'s
``anti-turbulence theorem'' states that the viscous dissipation of kinetic
energy is bounded from above by a bound that is independent of the Rayleigh
number. If $D(APE)$ increases with $R_a$ with no corresponding increase in
$D(KE)$, then one can infer that the dissipative mixing efficiency 
$\gamma_{mixing}$ must also increase with the Rayleigh number. 
Testing this hypothesis further will be an important future research goal.

\par

 As shown by \cite{Paparella2002}, the ``anti-turbulence theorem'' places a
stringent constraint on the magnitude of the viscous dissipation rate of
kinetic energy that can be sustained by buoyancy forcing alone. As a 
consequence, mechanical forcing appears critical to produce significant
amount of viscous dissipation of kinetic energy $D(KE)$. However, because
of the relatively low mixing efficiency of mechanically-driven mixing,
it is usually the case that the addition of mechanical forcing to 
horizontal convection results in relatively modest increase in $D(APE)$
compared to that in $D(KE)$. The overall consequence is to decrease the
mixing efficiency $\gamma_{mixing}$ of the system as compared to the 
purely buoyancy-driven case, as verified in the numerical experiments
considered here for which $G(KE)>0$. The decrease in mixing efficiency
resulting from mechanically stirring horizontal convection does not 
preclude an increase in $G(APE)$ however, as expected from the theoretical
formula, and verified in the numerical experiments. In our opinion,
this is the fundamental reason why the strength of the overturning circulation
in \cite{Whitehead2008}'s laboratory experiments appears to be greatly 
enhanced by the action of the stirring rod. Depending on the particular
circumstances, it appears possible for the mechanical forcing to contribute
directly to the observed overturning increase, in addition to the indirect
effect it has on the increase of $G(APE)$, as in the particular case of 
the clockwise mechanical forcing experiment. In the cases where the 
mechanical forcing acts anti-clockwise, however, only the increase in
$G(APE)$ appears to be directly responsible for the increase in the 
overturning. The latter case is likely to be the one pertaining to 
\cite{Whitehead2008}'s laboratory experiments, as it seems difficult to
see how the lateral motions of the stirring rod could contribute to the
vorticity source required to drive a thermally direct cell.

\par

  The present results make it hard to regard Eq. (\ref{general_constraint2})
as a constraint on the mechanical sources of stirring, as proposed by 
\cite{Munk1998}, given that $G(APE)$ is likely more sensitive to a
change in the mechanical forcing than $G(KE)$ to a change in the buoyancy
forcing, although this has not been directly verified. The other main 
difficulty in interpreting Eq. (\ref{general_constraint2}) as a constraint
on $G(KE)$ comes from the fact that the value of $\gamma_{mixing}$, far
from having the value of $0.2$ commonly affixed to it, appears to be 
extremely sensitive to the details of the mechanical and buoyancy forcing.
With the present experiments, values were found to range
between 0.6 and 1.78. The fact that buoyancy forcing is as important as the
mechanical stirring due to the wind and tides has important consequences for
the value of mixing efficiency one should use in the oceans. We note that if
one uses $\gamma_{mixing}=1$ in \cite{Munk1998}'s paper, the requirement on
the mechanical sources of stirring reduces drastically to 
$G(KE)=0.4\,{\rm TW}$, which is easily achieved by the wind alone.
Future research, therefore, should aim at understanding the physical
mechanisms controlling the value of mixing efficiency in mechanically- and
thermodynamically forced stratified fluids. We note that to date, 
mixing efficiency is traditionally studied in the context of freely decaying
turbulence, not in the context of forced/dissipated systems.
Interestingly, \cite{Deboer2008} suggest that coarse resolution numerical
ocean models behave as predicted by the present theory, as they found
no overturning
circulation cell in the total absence of wind forcing in the Atlantic ocean.
Another issue of interest are the consequences of mechanical control for
the multiple equilibria of the thermohaline circulation, recently discussed
by \cite{Johnson2007} and \cite{Nof2007}. Indeed, such a problem is usually
studied under the assumption that the turbulent mixing parameters can be
somehow regarded as fixed. The present results suggest, however, that 
the overall value of $\gamma_{mixing}$ might differ for different equilibria.

\section{Acknowledgements}

This work was created using the Tellus \LaTeXe\ class file.
This study was supported by the NERC funded RAPID programme. 
The author acknowledges comments by J. Whitehead, J. Nycander,
and an anonymous referee on an earlier draft 
which were helpful in improving the present manuscript.

\clearpage

\end{document}